\documentclass[preprint,review]{elsarticle} 

\usepackage[margin=1in]{geometry} 
\usepackage{amsmath,amssymb,amsfonts,amsthm}
\usepackage{stmaryrd,mathdots,bbm}
\usepackage{enumerate}
\usepackage{yfonts}
\usepackage{multicol}
\usepackage{ragged2e}
\usepackage{faktor}
\usepackage{pgfplots}
\usepackage{esint}
\usepackage{pinlabel}
\usepackage{graphicx}
\usepackage{subcaption}
\usepackage{tikz}
\usepackage{tikz-cd}
\usepackage[shortlabels]{enumitem}
\usepackage{longtable}
\usepackage{xparse}
\usepackage{mathrsfs}
\usepackage{caption}
\usepackage{lmodern}
\usepackage[utf8]{inputenc}
\usepackage[T1]{fontenc}
\usepackage{orcidlink}
\usepackage{hyperref}
\hypersetup{colorlinks,allcolors=blue}
\usepackage[capitalize]{cleveref}
\pgfplotsset{compat=1.18}

\newcommand{\Z}{\mathbb{Z}}

\NewDocumentCommand{\Hom}{mmg}{\ensuremath{\operatorname{Hom}_{\IfNoValueTF{#3}{}{#3}}(#1,#2)}}
\NewDocumentCommand{\Tor}{mmmg}{\ensuremath{\operatorname{Tor}_{#3}^{\IfNoValueTF{#4}{}{#4}}(#1,#2)}}
\NewDocumentCommand{\Ext}{mmmg}{\ensuremath{\operatorname{Ext}^{#3}_{\IfNoValueTF{#4}{}{#4}}(#1,#2)}}

\numberwithin{equation}{section}
\newtheorem{theorem}{Theorem}[section]

\theoremstyle{definition}

\journal{Chaos, Solitons \& Fractals}

\begin{document}

\begin{frontmatter}

\title{Validity condition of normal form transformation for the $\beta$-FPUT system}

\author[umich]{Boyang Wu\corref{cor1}\orcidlink{0000-0002-7937-2047}}
\ead{boyangwu@umich.edu}

\author[unito]{Miguel Onorato\orcidlink{0000-0001-9141-2147}}

\author[umich]{Zaher Hani\orcidlink{0000-0003-2195-0120}}

\author[umich-me]{Yulin Pan\orcidlink{0000-0002-7504-8645}}

\cortext[cor1]{Corresponding author}

\address[umich]{Department of Mathematics, University of Michigan, Ann Arbor, MI 48109, USA}
\address[unito]{Dipartimento di Fisica, Universit\`a di Torino, 10125 Torino, Italy}
\address[umich-me]{Department of Mechanical Engineering, University of Michigan, Ann Arbor, MI 48109, USA}

\begin{abstract}
In this work, we provide a validity condition for the normal form transformation to remove the non-resonant cubic terms in the $\beta$-FPUT system. We show that for a wave field with random phases, the normal form transformation is valid by dominant probability if $\beta \ll 1/N^{1+\epsilon}$, with $N$ the number of masses and $\epsilon$ an arbitrarily small constant. To obtain this condition, a bound is needed for a summation in the transformation equation, which we prove rigorously in the paper. The condition also suggests that the importance of the non-resonant terms in the evolution equation is governed by the parameter $\beta N$. We design numerical experiments to demonstrate that this is indeed the case for spectra at both thermal-equilibrium and out-of-equilibrium conditions. The methodology developed in this paper is applicable to other Hamiltonian systems where a normal form transformation needs to be applied.
\end{abstract}

\begin{keyword}
FPUT \sep wave turbulence \sep normal form \sep canonical transformation
\end{keyword}

\end{frontmatter}

\section{Introduction}
The Fermi-Pasta-Ulam-Tsingou (FPUT) chain is a model introduced in the 1950s to study the thermal equipartition in crystals \cite{EJS,Fermi1,Fermi2}. The model considers a chain of masses connected by springs with quadratic or cubic nonlinearities (so called $\alpha$-\hspace{0pt}FPUT or $\beta$-\hspace{0pt}FPUT respectively) in their restoring forces. Fermi expected the system to reach thermalization state (e.g. with equipartition of energy in all degrees of freedom) in a short time of evolution. However, when simulated on the computer MANIAC-\hspace{0pt}I, an unexpected recurrence was discovered, that the system returns to almost the exact initial condition (with energy only in large scales) after certain time of integration. Further interest and study in FPUT system led to the development of integrable theory, KdV solitons and Hamiltonian chaos \cite{schneider2000counter,zabusky1965interaction,gardner1967method}. To some extent, the FPUT recurrence is explained by the near-integrable property of the model, but the model was later found to indeed reach thermalization over a much longer time scale \cite{FPU,ponno2011two}, which required further analysis.

One important development to understand the thermalization of FPUT system was via the wave turbulence (WT) theory \cite{FPU,Spohn08,lukkarinen2008anomalous}. In \cite{FPU}, it was shown that the WT prediction of the $\alpha$-FPUT thermalization time scale agrees with their simulation results. The WT method was then further extended to study thermalization in $\beta$-FPUT and discrete nonlinear
Klein-Gordon (DNKG) models \cite{pistone2018universal}, again with agreement to simulations. The centerpiece of the WT-based argument is the derivation of a wave kinetic equation (WKE) that describes the evolution of the so-called wave action spectrum, whose evolution time scale provides the thermalization time scale.

An important procedure in deriving WKE for these systems is a normal form transformation, which is a near-identity transformation enabling one to remove non-resonant terms in the dynamical equation \cite{arnol2013mathematical,zakharov2012kolmogorov}. Roughly speaking, these transformations are mathematically possible because the frequency condition for non-resonant interactions cannot be satisfied, so the zero-divisor problem can be avoided. However, a non-vanishing divisor does not mean that the transformation remains bounded or near-identity, and this necessary property is usually not checked in many physics literature where a normal form transformation is applied. Physically, ignoring this validity condition has important consequences, since violation of the condition corresponds to the case of quasi-resonant interactions to be associated with significant energy transfer (when the non-vanishing divisor becomes small enough). Such interactions should not be removed in this case, which otherwise leads to deviation from the true dynamics.

Clearly, the application of normal form transformation should be accompanied by a validity condition in FPUT and other Hamiltonian systems \cite{pistone2018universal,bambusi2006metastability,Krasitskii_1994,janssen2009some,ZAKHAROV1985285}. To the authors’ knowledge, at least two works have been carried out in this direction. The first one is \cite{ganapa2023quasiperiodicity}, which attempts to establish a validity condition for the normal form transformation to remove quadratic terms in the $\alpha$-FPUT system. However, the analysis is based on an oversimplified system only containing a few modes. Their conclusion therefore cannot be generalized to cases with a broadband spectrum (e.g., the thermal equilibrium state). The second work is \cite{ZAKHAROV1999327}, which deals with normal form transformation in water waves of finite depth. The analysis is done for a broadband spectrum, but random phases of modes are not considered. Because of this treatment, significant cancellations in the estimation are ignored and the final result is not consistent with the most general setting of random waves.

In this paper, we establish a validity condition of the normal form transformation in a rigorous manner. Our analysis is conducted for the $\beta$-FPUT system to remove the four-wave non-resonant interactions, but the developed approach can be widely applied to other scenarios of Hamiltonian systems.  We prove rigorously that, for the above mentioned normal form transformation to stay near-identity, it is required that $\beta \ll 1/N^{1+\epsilon}$, with $N$ the number of masses and $\epsilon$ an arbitrarily small constant. This result suggests physically that the relative importance of non-resonant interactions to the dynamics is governed by the parameter $\beta N$ (with $\epsilon$ omitted). We numerically confirm this point by directly computing the fraction of non-resonant terms in the Hamiltonian of an $\beta$-FPUT system evolving from two initial conditions, one at thermal equilibrium and the other out-of-equilibrium. For both cases we find the fraction depends only on $\beta N$ as suggested by the theory.

\section{Review of normal form for $\beta$-FPUT model}
We first briefly review the formulation of the $\beta$-FPUT model and its associated normal form transformation following \cite{bustamante2019exact}. The $\beta$-FPUT model consists of $N$ identical masses $m$ connected by nonlinear springs with the elastic force $F=-\kappa\Delta q+\beta\Delta q^3$, where $\kappa,\beta\neq 0$ are spring elastic constants. The equation of motion for a $\beta$-FPUT chain can be expressed as (e.g. \cite{EJS,bustamante2019exact}):
\begin{align}
m\Ddot{q}_j&=\kappa\left[(q_{j+1}-q_j)-(q_j-q_{j-1})\right]+\beta\left[(q_{j+1}-q_j)^3-(q_j-q_{j-1})^3\right], \label{eq-1.1}
\end{align}
where $j=0,1,...,N-1$ and $q_j(t)$ represents the displacement from the equilibrium position of the mass $j$ with periodic boundary conditions $q_0=q_N$.

We consider a discrete Fourier transform
\begin{align}
Q_k&=\frac{1}{N}\sum_{j=0}^{N-1}q_je^{-i2\pi k_j/N},\mbox{ }P_k=\frac{1}{N}\sum_{j=0}^{N-1}p_je^{-i2\pi k_j/N},\label{eq-1.3}
\end{align}
with $p_j=m\dot{q}_j$ and introduce the normal modes $a_k(t)$ as
\begin{align}
a_k=\frac{\sqrt{2}}{2}\left[(m\omega_k)^{\frac{1}{2}}Q_k+i(m\omega_k)^{-\frac{1}{2}}P_k\right],  \mbox{ for }k=1,...,N-1. \label{eq-1.51}
\end{align}
The evolution equation \eqref{eq-1.1} then becomes
\begin{align}
i\dot{a}_{k_1}=&\omega_{k_1}{a}_{k_1}+\frac{\beta}{3}\sum_{k_2,k_3,k_4}\bigg[T_{1,2,3,4}a_{k_2}a_{k_3}a_{k_4}\delta(k_1-k_2-k_3-k_4)\notag\\
&+3T_{1,-2,3,4}a_{k_2}^*a_{k_3}a_{k_4}\delta(k_1+k_2-k_3-k_4)+3T_{1,-2,3,4}a_{k_2}a_{k_3}^*a_{k_4}^*\delta(k_1-k_2+k_3+k_4)\notag\\
&+T_{1,2,3,4}a_{k_2}^*a_{k_3}^*a_{k_4}^*\delta(k_1+k_2+k_3+k_4)\bigg],\label{eq-1.25}
\end{align}
where the frequency and interaction coefficients read
\begin{align}
    \omega_{k}&=\omega(k)=2\sqrt{\frac{\kappa}{m}}\left|\sin\left(\frac{\pi k}{N}\right)\right|,\\
    T_{1,2,3,4}&=- \frac{3}{4\kappa^2}\iota( k_2+k_3+k_4)\iota(k_2)\iota(k_3)\iota(k_4) \prod_{i=1}^4 \sqrt{\omega_{k_i}},
\end{align}
with $\iota(x):=\text{sgn} \sin \left(\frac{\pi x}{N}\right)$ a $2N$-periodic function. $\delta(\cdot)$ denotes the Kronecker delta with mod $N$ applied to its argument. Hereafter the mod $N$ condition is applied to all wavenumber conditions but sometimes not written explicitly for simplicity.

Among the four terms in \eqref{eq-1.25}, only the second one with $\delta(k_1+k_2-k_3-k_4)$ leads to resonant interactions. For the other three non-resonant terms, the corresponding frequency condition cannot be satisfied due to the the positivity and strict subadditivity of $\omega(k)$, as well explained in \cite{bustamante2019exact}. Therefore, we can remove these non-resonant terms at the current order via a normal form transformation written as
\begin{align}
a_{k_1} =& b_{k_1}+\frac{\beta}{3}\sum_{k_2,k_3,k_4}\bigg[A^{(1)}_{1,2,3,4}b_{k_2}b_{k_3}b_{k_4}\delta(k_1-k_2-k_3-k_4)\notag\\
&+3A^{(2)}_{1,2,3,4}b_{k_2}b_{k_3}^*b_{k_4}^*\delta(k_1-k_2+k_3+k_4)
+A^{(3)}_{1,2,3,4}b_{k_2}^*b_{k_3}^*b_{k_4}^*\delta(k_1+k_2+k_3+k_4)\bigg]. \label{eq-CT}
\end{align}
The goal here is that when \eqref{eq-CT} is substituted into \eqref{eq-1.25}, the resulting equation becomes
\begin{align}
i\dot{b}_{k_1}=&\omega_{k_1}{b}_{k_1}+\beta\sum_{k_1-k_2+k_3-k_4=0}T_{1,-2,3,4}b_{k_2}^*b_{k_3}b_{k_4}+O\left(\beta^2\right). \label{FPU}
\end{align}
In order for \eqref{FPU} to be achieved, one needs to choose the coefficients of $A^{(i)}_{1,2,3,4}$ accordingly as
\begin{align}
A^{(1)}_{1,2,3,4}=-\frac{T_{1,2,3,4}}{\omega_{k_1}-\omega_{k_2}-\omega_{k_3}-\omega_{k_4}},\label{A-1}\\
A^{(2)}_{1,2,3,4}=-\frac{T_{1,-2,3,4}}{\omega_{k_1}-\omega_{k_2}+\omega_{k_3}+\omega_{k_4}},\label{A-2}\\
A^{(3)}_{1,2,3,4}=-\frac{T_{1,2,3,4}}{\omega_{k_1}+\omega_{k_2}+\omega_{k_3}+\omega_{k_4}}\label{A-3}.
\end{align}


\section{Validity condition of the normal form transformation}
In order for the normal form transformation \eqref{eq-CT} to stay bounded and near-identity, we need 
\begin{align}
\frac{\beta}{3}\left|\sum_{k_2,k_3,k_4}A^{(1)}_{1,2,3,4}b_{k_2}b_{k_3}b_{k_4}\delta(k_1-k_2-k_3-k_4)\right| \ll \left|b_{k_1}\right|. \label{eq-req}
\end{align}
and similar conditions for other sums involving $A^{(2)}_{1,2,3,4}$ and $A^{(3)}_{1,2,3,4}$. A numerical study for similar conditions have been presented in \cite{comito2025role}. In addition, such a validity condition has also been formulated in \cite{ZAKHAROV1999327} for gravity waves in finite water depth. However, in further estimating the sums in the LHS of the condition, the random phases in $b_k$ are not considered in \cite{ZAKHAROV1999327}, leading to an overestimation. 

In order to account for the random phases in $b_k$, we consider a field with $b_k=\sqrt{\phi_k}\eta_k(\varrho)$, where $\phi_k$'s are deterministic amplitude of $O(1)$ and $\eta_k(\varrho)$'s are i.i.d. zero-mean complex random variables with unit variance, either as standard complex Gaussian or uniformly distributed on the unit circle. Under this condition, the sum in \eqref{eq-req} behaves like a random walk in a complex plane, for which cancellations due to walks in different directions must be considered. Intuitively this sum should be evaluated as the expected value $\mathbb{E}[|\sum \boldsymbol \cdot|]$ or $\{ \mathbb{E}[|\sum \boldsymbol \cdot|^2] \}^{1/2}$. Mathematically, this reasoning is guaranteed by the standard hyper-contractivity estimates as in \cite[Lemma 3.1]{2019}, which asserts that the probability of the sum
\begin{align}
\mathbb{P}\left(\left|\sum_{k_1-k_2-k_3-k_4=0}A^{(1)}_{1,2,3,4}b_{k_2}b_{k_3}b_{k_4}\right|\geq AM^{\frac{1}{2}}\right) \leq Ce^{-cA^{\frac{2}{3}}},
\end{align}
for some positive constants $A,C,c>0$, with
\begin{align}
M&=\mathbb{E}\left|\sum_{k_1-k_2-k_3-k_4=0}A^{(1)}_{1,2,3,4}b_{k_2}b_{k_3}b_{k_4}\right|^2.
\end{align}
This hyper-contractivity lemma simply states that with dominant probability, the sum in \eqref{eq-req} is estimated by $M^{1/2}$.

By further applying the Wick contraction theorem (see \cite[Theorem 5.1]{TextbookWT}), we obtain
\begin{align}
M = \sum_{k_1-k_2-k_3-k_4=0}\left|A^{(1)}_{1,2,3,4}\right|^2\phi_2\phi_3\phi_4.
\label{eq-Mest}
\end{align}

In the Appendix of the paper, we prove the following theorem.

\begin{theorem}\label{thm}
    Let $A^{(i)}_{1,2,3,4}$, $i=1,2,3$, be defined as in \eqref{A-1},\eqref{A-2}, and \eqref{A-3}, then for fixed $k_1 \in \mathbb{Z} \cap (0,N)$, we have:
    \begin{align}
        \sum_{\substack{k_1-k_2-k_3-k_4 = 0 \text{ } \textup{(mod $N$)}\\k_2,k_3,k_4 \in \mathbb{Z} \cap (0,N)}}\left|A^{(1)}_{1,2,3,4}\right|^2+\sum_{\substack{k_1-k_2+k_3+k_4 = 0 \text{ } \textup{(mod $N$)}\\k_2,k_3,k_4 \in \mathbb{Z} \cap (0,N)}}\left|A^{(2)}_{1,2,3,4}\right|^2\notag \\
        +\sum_{\substack{k_1+k_2+k_3+k_4 = 0 \text{ } \textup{(mod $N$)}\\k_2,k_3,k_4 \in \mathbb{Z} \cap (0,N)}}\left|A^{(3)}_{1,2,3,4}\right|^2 \lesssim N^2 \log N. \label{thm-1}
    \end{align} 
\end{theorem}

Combining \eqref{eq-req} (and the counterparts for other sums), \eqref{eq-Mest} and Theorem \ref{thm}, we conclude that the validity condition of the normal form transformation is
\begin{align}
\beta  \ll \frac{1}{N^{1+\epsilon}},
\label{eq-fincon}
\end{align} 
where $\epsilon$ is an arbitrarily small constant, resulting from the $\log N$ factor in \eqref{thm-1} for large $N$.

We note that to obtain \eqref{eq-fincon}, we have considered that $\phi_k \sim O(1)$ for all $k$. This means physically that the spectrum is broadband and does not deviate too much from the thermal-equilibrium state of equipartition of wave action. Of course, this is mathematically done for the convenience of cancellation of $\phi_i$'s in the condition. We will show numerically in next section that the condition derived under such a consideration does practically apply to spectra of different forms.

\section{Numerical Verification}
The condition \eqref{eq-fincon} suggests that the relative importance of the three non-resonant terms in \eqref{eq-1.25} to the dynamics is governed by the parameter $\beta N$ ($\epsilon$ omitted hereafter), which we aim to verify numerically in this section. We first note that the Hamiltonian for the original $\beta$-FPUT system is given by  
\begin{equation}
\begin{split}
&\frac{H}{N}=\sum_{k=1}^{N-1}\omega_k|a_k|^2+ \beta\sum_{k_1,k_2,k_3,k_4}
\bigg[
T_{1,2,3,4}(a_1^*a_2 a_3 a_4+c.c.) \delta_{1-2-3-4}
\\
&\qquad\qquad\qquad\qquad\qquad+
\frac{3}{2}T_{1,-2,3,4}a_1^*a_2^* a_3 a_4 \delta_{1+2-3-4}+\frac{1}{4}T_{1,2,3,4}(a_{1}^*a_{2}^*a_{3}^*a_{4}^*+c.c. )\delta_{1+2+3+4}
\bigg],
\end{split} \label{eq-Ham}
\end{equation}
where $c.c.$ denotes the complex conjugate of the previous term. The relative dynamical contribution from the non-resonant terms can be characterized by the ratio of non-resonant to resonant terms in \eqref{eq-Ham}. More specifically, we first define \begin{align}
S_{1}=&\sum_{k_1,k_2,k_3,k_4}
T_{1,2,3,4}(a_1^*a_2 a_3 a_4+c.c.) \delta_{1-2-3-4},\label{S-1}\\
S_{2}=&\sum_{k_1,k_2,k_3,k_4}\frac{3}{2}T_{1,-2,3,4}a_1^*a_2^* a_3 a_4 \delta_{1+2-3-4},\label{S-2}\\
S_{3}=&\sum_{k_1,k_2,k_3,k_4}\frac{1}{4}T_{1,2,3,4}(a_{1}^*a_{2}^*a_{3}^*a_{4}^*+c.c. )\delta_{1+2+3+4},\label{S-3}
\end{align} 
and then the ratio
\begin{align}
r = \left|\frac{\overline {S_1}+\overline {S_3}}{\overline {S_2}}\right|,
\label{eq-r}
\end{align}
where $\overline{\boldsymbol \cdot}$ represents average of the quantity (computed from time and ensemble average specified later). Our goal is to numerically compute $r$ as a function of $\beta$ and $N$ and show that $r$ depends on $\beta N$ irrespective of each factor, as theory suggests.

For the purpose mentioned above, we perform numerical simulations of \eqref{eq-1.1} using the sixth order symplectic integrator method \cite{yoshida1990construction}. We choose two sets of different initial conditions, one at thermal equilibrium (of energy equipartition):
\begin{align}
    a_k{(0)}=\sqrt{1/\omega_k}\cdot\exp(i\varphi_k), \label{ic-eq}
\end{align}
and one out of equilibrium:
\begin{align}
    a_k{(0)}=\sqrt{(1+\omega_k^2)/\omega_k}\cdot\exp(i\varphi_k), \label{ic-oeq}
\end{align}
with phases \(\varphi_k\) uniformly distributed from $[0,2\pi)$ for all $k\in \Z\cap(0,N)$. We can then recover $q_j(0),p_j(0)$ from \eqref{eq-1.3} and \eqref{eq-1.51}, as the initial condition for \eqref{eq-1.1}. We set $m = 1, \kappa = 1$, so that the fundamental period $T_f=\frac{2\pi}{\omega(k=1)}\approx N$. We set the time step $h=0.01$ and simulate the system up to $t_{\text{max}}=10T_f$. The average in \eqref{eq-r} is computed from time average over $[5T_f, 10T_f]$, then ensemble average over 5 ensembles. 

\begin{center}
    \includegraphics[width=1\linewidth]{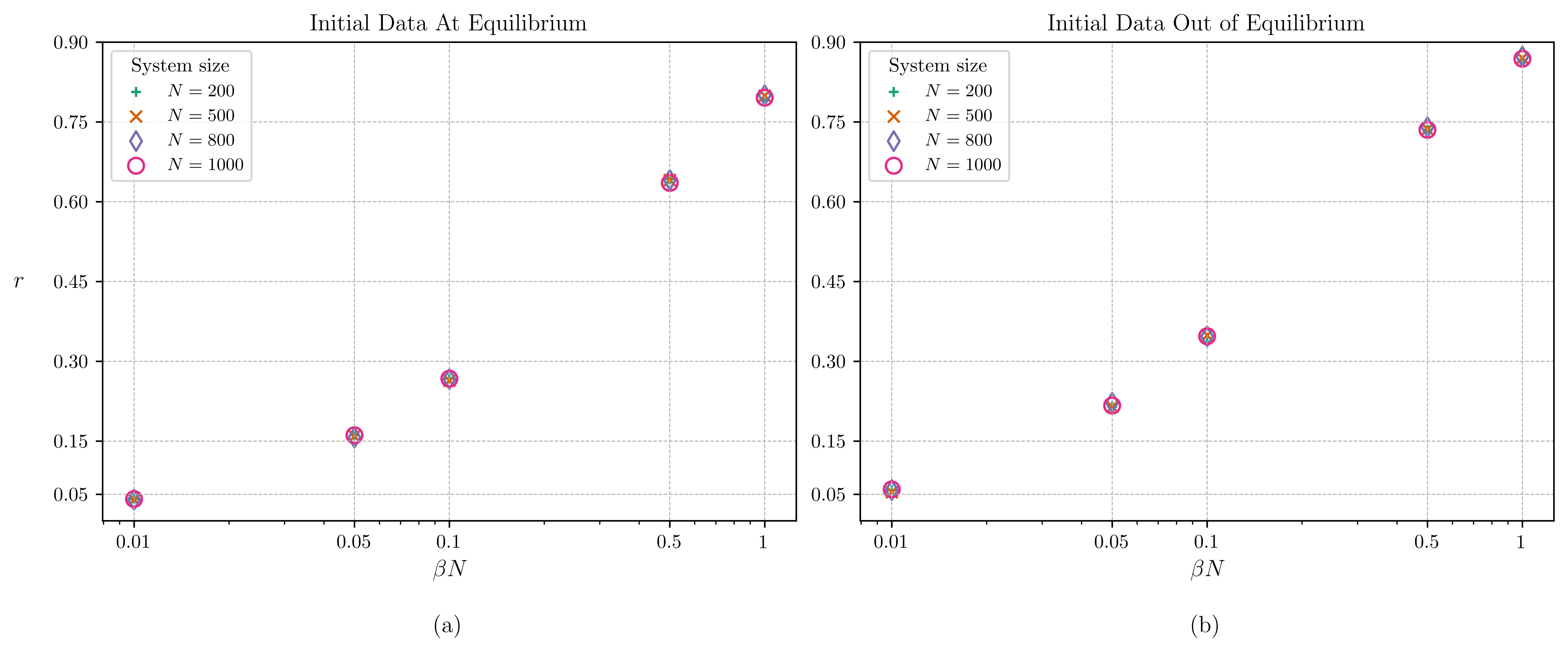}
    \captionof{figure}{Ratio \(r\) as a function of $\beta N$ for $N = 200$,  $500$,  $800$, and $1000$, with initial conditions (a) at thermal equilibrium with \eqref{ic-eq} and (b) out of thermal equilibrium with \eqref{ic-oeq}.}
    \label{fig:ratio-sum}
\end{center}

Figure \ref{fig:ratio-sum} plots the computed values of $r$ from the two initial conditions \eqref{ic-eq} and \eqref{ic-oeq} for different values of $\beta$ and $N$. In particular, we choose $N = 200$,  $500$,  $800$, and $1000$, and for each $N$ corresponding values of $\beta$ to make $\beta N=1$,  $0.5$, $0.1$, $0.05$, and $0.01$. We see from figure \ref{fig:ratio-sum} that all results of $r$ collapse for the same value of $\beta N$ irrespective of the value of $N$, which is true for both initial conditions. This means that the relative contributions from non-resonant terms are indeed controlled by the parameter $\beta N$ as indicated by the theory. In addition, we see that for large values of $\beta N$, say $\beta N=1$, the contribution from non-resonant terms can rise up to $O(80\%)$ for both initial conditions. Clearly for these cases, ignoring the non-resonant terms (i.e., applying the normal form transformation inconsistently) lead to excessive errors to the dynamics of the system.

\section{Conclusion}
We have derived a rigorous upper bound on the nonlinearity parameter $\beta$ as $\beta\ll 1/N^{1+\epsilon}$, that guarantees the normal form transformation for the $\beta$‑FPUT system to remain bounded and near‑identity. This validity condition suggests a parameter $\beta N$, which controls the relative importance of the non-resonant cubic terms in the dynamical evolution equation. We further design numerical experiments to verify this condition by applying the a sixth-order symplectic integrator method to the dynamical equation. We show that for both initial conditions at thermal equilibrium and out-of-equilibrium, the relative contribution of non-resonant terms in the Hamiltonian is indeed governed by $\beta N$ as indicated by the theory. Beyond a clarification of the validity condition, our work provides a practical criterion to select $\beta$ and $N$ with awareness of the effect of non-resonant terms for future studies of FPUT dynamics. The methodology to achieve so can also be extended to normal form transformations in other nonlinear Hamiltonian systems. Furthermore, the present results provide a rigorous justification for employing normal-form transformations to extend the derivation of the wave kinetic equation from the reduced evolution equation of the $\beta$-FPUT system \cite{RDFPUT} to the full $\beta$-FPUT dynamics.

{\bf Acknowledgments}
This research was supported by Simons Collaboration on Wave Turbulence, Grant No. 617006. M.O. is also supported by INFN (MMNLP and FieldTurb). Computational resources and services were provided by Advanced Research Computing (ARC) at the University of Michigan, including the Great Lakes high-performance computing cluster.

\appendix
\section{Proof of Theorem \ref{thm}}
To prove Theorem \ref{thm}, we separate the sums in the canonical transformation \eqref{eq-CT} into three cases with respect to the $\delta(\cdot)$ functions and find upper bounds for the sums. Note that all $k's$ are positive as defined in \eqref{eq-1.51}.\\
\underline{Case 1}: $k_1-k_2-k_3-k_4 \equiv 0 \text{ } \text{(mod } N)$. Then we may have three possibilities.
    \begin{enumerate}
        \item $k_1-k_2-k_3-k_4=0$, then we have:
        \begin{align}
        &\omega_{k_1}-\omega_{k_2}-\omega_{k_3}-\omega_{k_4}=2\left[\sin{\left(\frac{\pi k_1}{N}\right)}-\sin{\left(\frac{\pi k_2}{N}\right)}-\sin{\left(\frac{\pi k_3}{N}\right)}-\sin{\left(\frac{\pi k_4}{N}\right)}\right]\notag\\
        &=4\left[\sin{\left(\frac{\pi (k_1-k_2)}{2N}\right)}\cos{\left(\frac{\pi (k_1+k_2)}{2N}\right)}-\sin{\left(\frac{\pi (k_3+k_4)}{2N}\right)}\cos{\left(\frac{\pi (k_3-k_4)}{2N}\right)}\right]\notag\\
        &=4\sin{\left(\frac{\pi (k_1-k_2)}{2N}\right)}\left[\cos{\left(\frac{\pi (k_1+k_2)}{2N}\right)}-\cos{\left(\frac{\pi (k_3-k_4)}{2N}\right)}\right]\notag\\
        &=8\sin{\left(\frac{\pi (k_1-k_2)}{2N}\right)}\sin{\left(\frac{\pi (k_1+k_2+k_3-k_4)}{4N}\right)}\sin{\left(\frac{\pi (k_3-k_4-k_1-k_2)}{4N}\right)}\notag\\
        &=8\sin{\left(\frac{\pi (k_1-k_2)}{2N}\right)}\sin{\left(\frac{\pi (k_1-k_4)}{2N}\right)}\sin{\left(\frac{\pi (k_3-k_1)}{2N}\right)}.
    \end{align}
        We must have $k_2,k_3,k_4<k_1$ with at least one of $k_i \sim k_1$ in the restriction of $k_1-k_2-k_3-k_4=0$. Then for fixed $k_1$, there are $O(k_1^2)$ choices for $(k_2,k_3,k_4)$. We choose a constant $C =\frac{1}{3}$. Without loss of generality suppose $k_2 \geq Ck_1$, then $k_1-k_2=k_3+k_4 \geq \max(k_3,k_4)$, $k_1-k_3=k_2+k_4\geq Ck_1$ and similarly $k_1-k_4\geq Ck_1$. Then we have:
        \begin{align}
        \left|A^{(1)}\right|^2 \lesssim \frac{\left(\frac{ k_1}{N}\right)^2\left(\frac{\max(k_3,k_4)}{N}\right)^2}{\left(\frac{\max(k_3,k_4)}{N}\right)^2\left(\frac{Ck_1}{N}\right)^4} \lesssim \frac{N^2}{k_1^2}.
        \end{align}
        Then we know that:
        \begin{align}
            \sum_{k_1-k_2-k_3-k_4=0}\left|A^{(1)}\right|^2 \lesssim \frac{N^2}{k_1^2}\cdot k_1^2 = N^2 
        \end{align}
    
        \item $k_1-k_2-k_3-k_4=-2N$, then we can do a change of variable: $k_1' = N-k_1, k_2' = N-k_2, k_3'=N-k_3, k_4'=N-k_4$, we then get:
        $k_1'-k_2'-k_3'-k_4'=(N-k_1)-(N-k_2)-(N-k_3)-(N-k_4)=-k_1+k_2+k_3+k_4-2N = 0$. And we still have $\sin\left(\frac{\pi k_i}{N}\right)=\sin\left(\frac{\pi k'_i}{N}\right)$. Then applying the result from the last case, we can get:
        \begin{align}
            \sum_{k_1-k_2-k_3-k_4=-2N}\left|A^{(1)}\right|^2 \lesssim N^2
        \end{align}

        \item $k_1-k_2-k_3-k_4=-N$, without loss of generality suppose $\frac{N|h|}{\pi}=|k_1-k_4|\leq|k_1-k_2|,|k_1-k_3|$, $k_2 \leq k_3$, then we can write $\frac{\pi k_1}{N}=y,\frac{\pi k_4}{N}=y+h,\frac{\pi k_2}{N}=x,\frac{\pi k_3}{N}=\pi-x-h$, which gives:
        \[\frac{\pi}{N}(k_1-k_2-k_3-k_4) = y-x-(\pi-x-h)-(y+h)=-\pi\]
        Then now we have:
        \begin{align}
            &\left|\sin{\left(\frac{\pi k_1}{N}\right)}-\sin{\left(\frac{\pi k_2}{N}\right)}-\sin{\left(\frac{\pi k_3}{N}\right)}-\sin{\left(\frac{\pi k_4}{N}\right)}\right|\notag\\
            =&\left|\sin(y)-\sin(y+h)-\sin(x)-\sin(\pi-x-h)\right|\notag\\
            \geq&\Big||\sin(y)-\sin(y+h)|-|\sin(x)+\sin(x+h)|\Big|\notag\\
            =&2\Bigg|\left|\cos\left(y+\frac{h}{2}\right)\sin\left(\frac{h}{2}\right)\right|-\left|\sin\left(x+\frac{h}{2}\right)\cos\left(\frac{h}{2}\right)\right|\Bigg| \label{case2}
        \end{align}
        Since we assume $k_2\leq k_3$, we have: $x\leq\pi-x-h<\pi$, which then gives $\frac{|h|}{2}<x+\frac{h}{2}\leq \frac{\pi}{2}$. It means that:
        \[\sin\left(\frac{|h|}{2}\right)<\sin\left(x+\frac{h}{2}\right).\]
        On the other hand, if $y\geq\frac{\pi}{2}$, then we must have $\frac{\pi}{2}\leq y-|h|\leq\pi$ or $\frac{\pi}{2}\leq y+|h|\leq\pi$. Similarly, if $y<\frac{\pi}{2}$, then we must have $0\leq y-|h|<\frac{\pi}{2}$ or $0\leq y+|h|\leq\frac{\pi}{2}$. Otherwise either $k_1> \frac{N}{2},k_2,k_3,k_4 \leq \frac{N}{2}$ or $k_1\leq \frac{N}{2}, k_2,k_3,k_4 > \frac{N}{2}$, which neither give $k_1-k_2-k_3-k_4=-N$. It then means that $|h|<\frac{\pi}{2}$ and
        \[0\leq\left|\cos\left(y+\frac{h}{2}\right)\right|\leq\left|\cos\left(\frac{h}{2}\right)\right|.\]
        Then we further get:
        \begin{align}
            \eqref{case2} \gtrsim \cos\left(\frac{h}{2}\right)\left|\sin\left(x+\frac{h}{2}\right)-\sin\left(\frac{|h|}{2}\right)\right|,
        \end{align}
        \begin{enumerate}[label=(\roman*)]
            \item If $h> 0$, then we have $\frac{x+h}{2}=\frac{1}{2}(x+\frac{h}{2})+\frac{h}{4}\leq \frac{3}{8}\pi$, and:
            \[\eqref{case2} \gtrsim \cos\left(\frac{x+h}{2}\right) \sin\left(\frac{x}{2}\right) \gtrsim \sin\left(\frac{x}{2}\right),\]
            then we can bound $\left|A^{(1)}\right|^2$ as:
            \begin{align}
                \left|A^{(1)}\right|^2 &\lesssim \frac{\sin(x)\sin(y)\sin(y+h)}{\sin^2\left(\frac{x}{2}\right)} \lesssim \frac{\sin(y)\sin(y+h)}{\sin\left(\frac{x}{2}\right)}.
            \end{align}
            In this case, since $x+\frac{h}{2}\leq \frac{\pi}{2}$, then $x<\frac{\pi}{2}$. We either have: $\pi-x-h\leq y-h$, which then gives $\frac{\pi}{2}< \pi-x\leq y$ and 
            $\frac{\sin(y)}{\sin(\frac{x}{2})}=\frac{2\sin(y)\cos(\frac{x}{2})}{\sin(x)}\lesssim 1$, or have: $\frac{\pi}{2}>x\geq y+h$, and $\frac{\sin(y+h)}{\sin(\frac{x}{2})}=\frac{2\sin(y+h)\cos(\frac{x}{2})}{\sin(x)}\lesssim 1$. Hence we can get $\left|A^{(1)}\right|^2 \lesssim 1$  for $h>0$.
            
            \item If $h = 0$, then
            \begin{align}
                \left|A^{(1)}\right|^2 &\lesssim 1.
            \end{align}
            \item If $h < 0$, then we have $\frac{x}{2}\leq\frac{\pi}{4}+\frac{|h|}{4}\leq \frac{3}{8}\pi$, and:
        \[\eqref{case2} \gtrsim \sin\left(\frac{x+h}{2}\right) \cos\left(\frac{x}{2}\right)\gtrsim \sin\left(\frac{x+h}{2}\right) ,\]
        then we can bound $\left|A^{(1)}\right|^2$ as:
        \begin{align}
            \left|A^{(1)}\right|^2 &\lesssim \frac{\sin(x+h)\sin(y)\sin(y+h)}{\sin^2\left(\frac{x+h}{2}\right)} \lesssim \frac{\sin(y)\sin(y+h)}{\sin\left(\frac{x+h}{2}\right)},
        \end{align}
         Similarly, since $x+\frac{h}{2}\leq \frac{\pi}{2}$, then $x+h<\frac{\pi}{2}$. We either have: $\frac{\pi}{2}<\pi-x-h\leq y-|h|=y+h$, which then gives $\frac{\sin(y+h)}{\sin(\frac{x+h}{2})}\lesssim 1$, or have: $x\geq y+|h|$, i.e. $\frac{\pi}{2}> x+h\geq y$ and $\frac{\sin(y)}{\sin(\frac{x+h}{2})}\lesssim 1$. Hence we can get $\left|A^{(1)}\right|^2 \lesssim 1$ for $h<0$.
        \end{enumerate}
    
        Let $\Z_N:=N^{-1}\Z$. Thus we can get the estimate for the sum:
        \begin{align}
            \sum_{k_1-k_2-k_3-k_4=-N}\left|A^{(1)}\right|^2 &\lesssim \sum_{h\in \mathbb{Z}_N \cap (0,1)} \sum_{x\in \mathbb{Z}_N \cap (0,1)} 1\lesssim N^2.
        \end{align}
    \end{enumerate}
    
    From the above cases, we have:
    \begin{align}
        \sum_{k_1-k_2-k_3-k_4 \equiv 0 \text{ } \text{(mod } N)}\left|A^{(1)}\right|^2 \lesssim N^2.
    \end{align}
\underline{Case 2}: $k_1-k_2+k_3+k_4 \equiv 0 \text{ } \text{(mod } N)$. Then we again have three possibilities.
    \begin{enumerate}
        \item $k_1-k_2+k_3+k_4 = 0$. Similarly as Case 1(1), we have:
        \begin{align}
        &\omega_{k_1}-\omega_{k_2}+\omega_{k_3}+\omega_{k_4}\notag\\
        &=8\sin{\left(\frac{\pi (k_2-k_1)}{2N}\right)}\sin{\left(\frac{\pi (k_2-k_3)}{2N}\right)}\sin{\left(\frac{\pi (k_2-k_4)}{2N}\right)}
        \end{align}
        And $k_1,k_3,k_4<k_2$. Similarly we choose a constant $C<\frac{1}{3}$. Without loss of generality, suppose $k_1 \geq Ck_2$, then $k_2-k_1 = k_3+k_4$, $k_2-k_4=k_1+k_3\geq k_1\geq Ck_2$, and similarly $k_2-k_3\geq Ck_2$, then we have
            \begin{align}
            \left|A^{(2)}\right|^2 \lesssim \frac{\left(\frac{1}{N}\right)^4k_1k_2k_3k_4}{\left(\frac{C k_2}{N}\right)^4\left(\frac{k_3+k_4}{N}\right)^2} \lesssim \frac{N^2}{k_2^2}.
            \end{align}
        Then we can calculate the sum:
        \begin{align}
            \sum_{k_1-k_2+k_3+k_4=0}\left|A^{(2)}\right|^2 \lesssim \sum_{k_2=1}^{N-1}\sum_{k_3=1}^{k_2}\frac{N^2}{k_2^2} \lesssim N^2\log N
        \end{align}
        \item $k_1-k_2+k_3+k_4 = 2N$. Then similarly as Case 1(2), we can change of variables: $k_i'=N-k_i$ for $i\in \{1,2,3,4\}$. Then we can obtain the same result as (1):
        \begin{align}
            \sum_{k_1-k_2+k_3+k_4=2N}\left|A^{(2)}\right|^2 \lesssim  N^2\log N.
        \end{align}
        \item $k_1-k_2+k_3+k_4 = N$. We apply the same technique as in Case 1(3) and we can conclude:
        \begin{align}
            \sum_{k_1-k_2+k_3+k_4=N}\left|A^{(2)}\right|^2 \lesssim  N^2.
        \end{align}
    \end{enumerate}
    Thus we have:
    \begin{align}
        \sum_{k_1-k_2+k_3+k_4 \equiv 0 \text{ } \text{(mod } N)}\left|A^{(2)}\right|^2 \lesssim N^2  \log N.
    \end{align}
\underline{Case 3}: $k_1+k_2+k_3+k_4 \equiv 0 \text{ } \text{(mod } N)$

    Let $M:=\max\left(\sin{\left(\frac{\pi k_1}{N}\right)},\sin{\left(\frac{\pi k_2}{N}\right)},\sin{\left(\frac{\pi k_3}{N}\right)},\sin{\left(\frac{\pi k_4}{N}\right)}\right)\leq 1$, and 
    
    $\qquad m:=\min\left(\sin{\left(\frac{\pi k_1}{N}\right)},\sin{\left(\frac{\pi k_2}{N}\right)},\sin{\left(\frac{\pi k_3}{N}\right)},\sin{\left(\frac{\pi k_4}{N}\right)}\right)$.

    Then we know that:
    \begin{align}
        \left|A^{(3)}\right|^2 \lesssim \frac{mM^3}{M^2} \lesssim 1
    \end{align}
    
    Thus we have:
    \begin{align}
        \sum_{k_1+k_2+k_3+k_4 \equiv 0 \text{ } \text{(mod } N)}\left|A^{(3)}\right|^2 \lesssim N^2.
    \end{align}
Therefore we have proved Theorem \ref{thm}.

\bibliographystyle{elsarticle-num} 
\bibliography{ref}

\end{document}